\documentclass[%
preprint,
showpacs,
amsmath,amssymb,
aps,prf,
]{revtex4-1}
\usepackage{booktabs}

\usepackage{tabularx}  

\usepackage{float}
\usepackage{graphicx}

\usepackage{bm}

\usepackage{subfigure}
\usepackage{color}
\usepackage{tikz}
\usepackage{CJKutf8}
\newcommand*{\circled}[1]{\lower.7ex\hbox{\tikz\draw (0pt, 0pt)%
    circle (.5em) node {\makebox[0em][c]{\small #1}};}}

\usepackage[colorlinks,citecolor = blue, linkcolor=black,hyperindex,CJKbookmarks]{hyperref}

\begin{document}
\title{Droplet dissolution driven by emerging thermal gradients and Marangoni flow}

\author{Binglin Zeng$^{1,2,3}$}
\author{Yuliang Wang$^{2,4}$}\email{wangyuliang@buaa.edu.cn}
\author{Christian Diddens$^{1}$}
\author{Harold J. W. Zandvliet$^{3}$}\email{h.j.w.zandvliet@utwente.nl}
\author{Detlef Lohse$^{1,5}$}\email{d.lohse@utwente.nl}

\affiliation{$^1$Physics of Fluids Group and Max Planck Center for Complex Fluid Dynamics, MESA+ Institute and J. M. Burgers Centre for Fluid Dynamics, University of Twente, P.O. Box 217, 7500AE Enschede, The Netherlands\\
$^2$School of Mechanical Engineering and Automation, Beijing Advanced Innovation Center for Biomedical Engineering, Beihang University, 37 Xueyuan Rd, Haidian District, Beijing, China\\
$^3$Physics of Interfaces and Nanomaterials, MESA+ Institute, University of Twente, P.O. Box 217, 7500 AE Enschede, The Netherlands\\
$^4$Ningbo Institute of Technology, Beihang University, 37 Xueyuan Rd, Haidian District, Beijing, China\\
$^5$Max Planck Institute for Dynamics and Self-Organization, Am Fassberg 17, 37077 G\"ottingen, Germany}

\date{\today}

\begin{abstract}
The lifetime $\tau$ of an isothermal
 and
 purely diffusively dissolving droplet in a host liquid scales as $\tau \sim R_0^2$
with its initial radius $R_0$ [Langmuir, Phys. Rev. 12, 368 (1919)].
For a droplet dissolving  due to natural convection driven by density differences, its  lifetime scales as
$\tau\sim R_0^{5/4}$ [Dietrich {\it et al.}, J. Fluid Mech.\ 794, 45 (2016)].
In this paper we experimentally find and theoretically derive
yet another droplet dissolution behavior,
resulting in
$\tau \sim R_0^4$. It occurs when the dissolution dynamics is controlled by local heating of the liquid,
leading to a modified  solubility and
a  thermal Marangoni flow around the droplet. The thermal gradient is achieved by
plasmonic heating of
a gold nanoparticle decorated sample surface, on which  a sessile
water droplet immersed in water-saturated 1-butanol solution is sitting.
The resulting off-wall thermal Marangoni flow
and the temperature dependence of the solubility
 determine  the droplet dissolution rate,
resulting in a shrinkage $R(t) \sim  (\tau -t )^{1/4}$ of the droplet radius and thus in $\tau \sim R_0^{4}$.



\end{abstract}


\maketitle
\section{Introduction}
Droplet dissolution in another liquid  is a process of utmost importance in the process technology. It is
 ubiquitous in a wide range  of important applications, such as
in the food and  cosmetic industry \cite{gupta2016}, for
drug delivery \cite{chou2015micromachines,wang2009jcis}, in micro- and nanoextraction
\cite{jain2011,lohse2016jfm,lohse2020nat},
micro-fabrication \cite{Lu2017Nanoscale,um2014jmcb},
autochemotaxis of droplets
\cite{jin2017pnas,liebchen2018},
etc.

On small scales,  the droplet dissolution process  is purely diffusive, controlled by the
 concentration gradient outside the droplet, normal to  its surface. The process  is mathematically equivalent to
 droplet evaporation in an ambient gas or bubble dissolution in an ambient liquid, see the review on
 nanodroplets and nanobubbles by Lohse and Zhang \cite{lohse2015rmp},
 prior reviews on droplet evaporation \cite{cazabat2010,erbil2012},
  or  classical papers by
 Langmuir \cite{langmuir1918},
 Epstein and Plesset
 \cite{epstein1950},  Deegan \cite{deegan1997},  Popov \cite{popov2005}, or in the context of evaporating aerosols,
 Wells \cite{wells1934}.
 For such droplets, when in thermal equilibrium with their surroundings,
 the radius $R$ decreases in time with a square root behavior $R(t) \sim (\tau - t)^{1/2}$,
 implying that the lifetime $\tau$ of the droplet is quadratic with its initial radius,  $\tau\sim R_0^2$, the
 so-called
 ``D$^2$ law'', where ``D'' stands for the droplet diameter (twice the radius $R_0$).

For larger scales, however,
natural convection  can become important for the dissolution process.
This holds for both  bubbles \cite{enriquez2014} and  droplets \cite{dietrich2016a,chong2020}, and
without an external flow, i.e., natural convection. The origin of the natural convection
is a  density difference emerging through the dissolution process
(local depletion or enhancement of one heavier or lighter species), leading to buoyant forces,
 locally driving
the natural convection and considerably enhancing the dissolution rate.
 Dietrich {\it et al.} \cite{dietrich2016a} and Chong {\it et al.} \cite{chong2020}
could experimentally, numerically, and theoretically show that under such conditions the droplet radius
of long-chain alcohols dissolving in water
shrinks as $R(t) \sim (\tau - t)^{4/5}$, leading to a droplet lifetime $\tau \sim R_0^{5/4}$.

However, the droplet dissolution process can not only induce body forces such as buoyancy, but also
surface forces, such as Marangoni forces  \cite{scriven1960,lohse2020nat,dewit2020}. For example,
Tan {\it et al.}  \cite{tan2019}
showed  that the solutal Marangoni flow at the interface
of  a binary water-ethanol droplet  dissolving in anethole oil
governs the   drop dissolution process. Escobar {\it et al.} \cite{escobar2020}
observed a dramatically enhanced and temporally
non-monotonic droplet dissolution rate, which was triggered by solutal Marangoni
flow,  in the peculiar geometry of a water-immersed
sessile long-chain alcohol droplet with an entrapped bubble.

Not only emerging concentration gradients
can induce body or surface forces acting back on the droplet dissolution process itself,
but also thermal gradients can do so.
When the droplet is immersed in another liquid, a temperature gradient not only induces thermal Marangoni flow, but also changes the solubility and hence the saturation concentration of the host liquid.
This leads to a
complex dissolution dynamics of the droplet, which in this paper
 we want to explore under controlled conditions.

The system we pick is a
 sessile water droplet   sitting on a gold nanoparticle decorated surface and immersed
 in water-saturated 1-butanol.
 Thanks to the    plasmonic  effect of the gold nanoparticles,
 the area under the droplet can locally be heated
 \cite{baffou2013,baffou2014,wang2018}:
 Upon irradiation with
  a continuous wave laser at the  plasmonic resonance of the gold nanoparticles,
  a huge amount of
  thermal energy is locally deposited under the sessile droplet, leading to a strong temperature gradient.
  Under these conditions we experimentally
 find  that the droplet dissolution time scales with the initial droplet radius
 as $\tau \sim R_0^4$. We then theoretically explain this new scaling behavior, based on an
 analysis of the thermal Marangoni
 forces, the thermal diffusion equation,
  and of the  temperature dependent solubility of water in 1-butanol.

 The paper is organized as follows: We briefly explain the experimental setup and the methods (section II).
 We then  report our experimental findings (section III), which are explained in section IV. The paper ends with
 conclusions and an outlook (section V).

\section{Experimental setup and methods}
In the experiment, the gold nanoparticle decorated substrate was placed in a quartz glass cuvette (10$\times$10$\times$45 mm) filled with \textcolor{black}{a} water 1-butanol mixture. \textcolor{black}{Since the 1-butanol mixture was oversaturated with water}, micro-sized water droplets nucleated on the substrate.
We work in the limit of low sessile droplet area density so that no droplet-droplet  interactions can occur.
A continuous wave laser (Cobolt Samba) of 532 nm wavelength was used to irradiate the \textcolor{black}{sample beneath the droplet from the bottom side}. The laser power was controlled by using a half-wave plate and a polarizer and measured by a photodiode power sensor (S130C, ThorLabs). Two high-speed cameras were installed in the setup to monitor the dynamics of the dissolving droplet. One (Photron SA7) was equipped with a 5$\times$ long working distance objective (LMPLFLN, Olympus) for bottom view imaging, and the other one (Photron SAZ) with a 10$\times$ or 20$\times$ long working distance objective and operated at 1 kfps for fast imaging (For details in sample preparation and the experimental setup, please refer to our previous study in Ref. \cite{wang2018}).

\section{Experimental results}

\begin{figure}[htpb]
\includegraphics[width=1\textwidth]{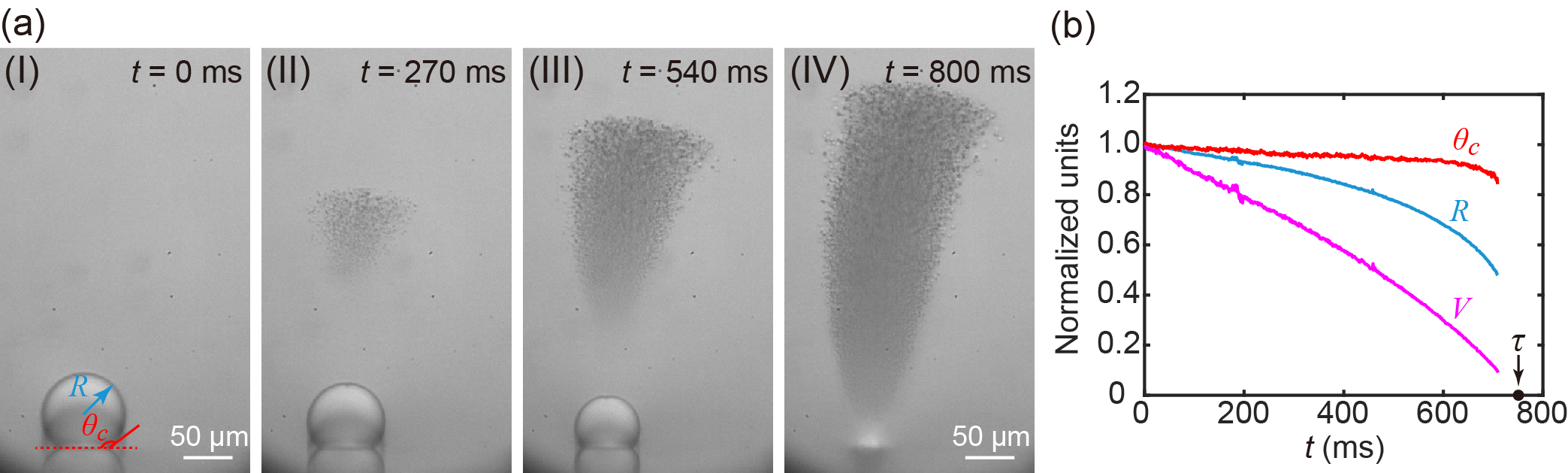}
\caption{(a) Snapshots of a dissolving water droplet in water-saturated 1-butanol at different instants of time under laser power $P_\ell$ = 40 mW. (b) Evolution of volume $V$, radius $R$, and contact angle $\theta_c$ in time. All parameters have been normalized by their initial values: $V_{0}$ = 0.32 nL, $R_0$ = 44 $\mu$m, and $\theta_0 \approx$ 130$^\circ$, \textcolor{black}{respectively}.}
\label{process}
\end{figure}

Sequential snapshots of a typical dissolving water droplet in water-saturated 1-butanol solution under laser irradiation are shown in Fig. \ref{process}a. \textcolor{black}{Immediately} after laser irradiation, the droplet starts to shrink (Fig. \ref{process}a-I). At 270 ms, a plume appears and moves off-wall away from
 the droplet (Fig. \ref{process}a-II). With time, the plume grows (Fig. \ref{process}a-III) until the droplet is completely dissolved at 800 ms (Fig. \ref{process}a-IV).
Figure \ref{process}b displays the volume $V$, radius $R$, and contact angle $\theta_c$ of the droplet as a function of time $t$ during the dissolution process. All values are normalized to their values at $t$ = 0, which is the moment the laser was turned on.
In the figure, $\tau$ refers to the moment when the droplet is completely dissolved, i.e.,
$\tau$ is the
 lifetime of the droplet. It is found that the contact angle $\theta_c$ of the droplet remains almost constant and only slightly decreases with time \textcolor{black}{at} the end of the dissolution process, which reflects
 dissolution in the  constant contact angle mode \cite{lohse2015rmp}, presumably due to pinning effects.
The  details of the whole dissolution process can be seen in  Movie S1.

\begin{figure}[htpb]
\includegraphics[width=0.5\textwidth]{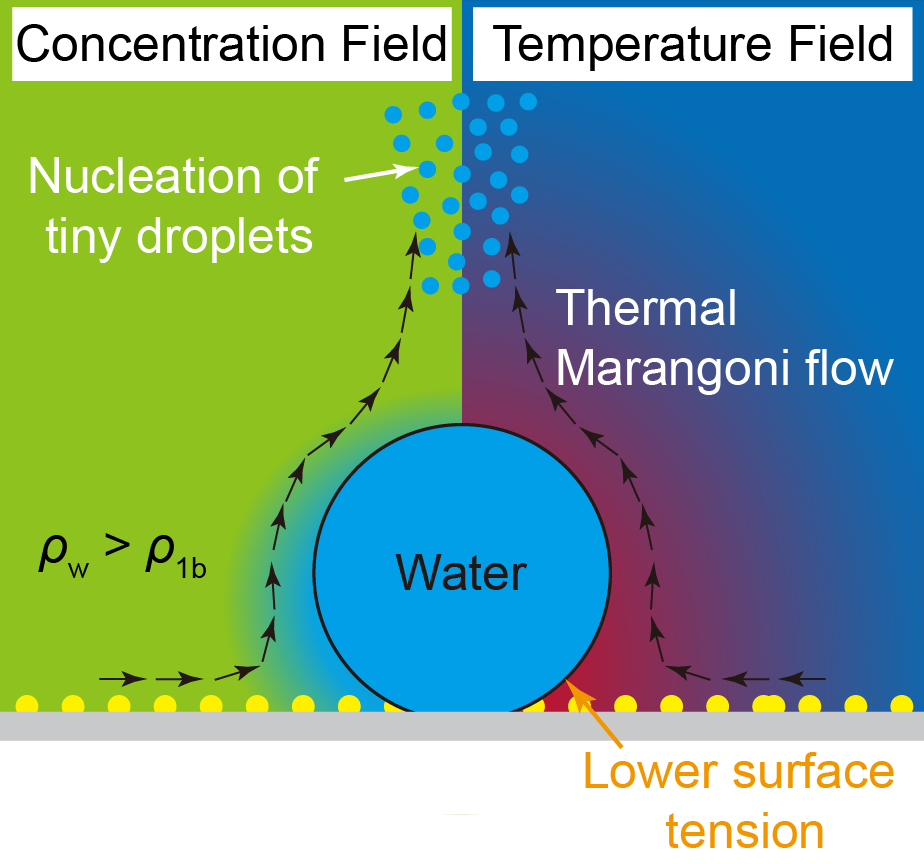}
\caption{Schematic diagram of the plume formation above a dissolving water droplet in the surrounding
water-saturated 1-butanol. The plasmonic effect induced temperature field
leads to  \textcolor{black}{an} increased water solubility and \textcolor{black}{a} decreased surface tension of 1-butanol solution \textcolor{black}{near} the bottom of the water droplet. As a result, an
off-wall  thermal Marangoni flow is generated, \textcolor{black}{transporting} the hot 1-butanol solution with a \textcolor{black}{high} water concentration to the cold region above the droplet, leading to the nucleation
of a water droplet plume. As explained in the main text, gravity does not play any role here; the flow
is solely driven by thermal Marangoni forces.}
\label{schematic}
\end{figure}

The dissolution of the water droplet is due to the increasing solubility of water in 1-butanol with increasing temperature \cite{Stephenson1986}.
As sketched  in Fig. \ref{schematic}, the liquid around the droplet is heated up because of the plasmonic effect. \textcolor{black}{The flow of water from the droplet into the 1-butanol solution implies
 a shrinkage of the water droplet.}
Besides the increased solubility of water in 1-butanol, the thermal Marangoni flow contributes to the droplet dissolution as well.
\textcolor{black}{The temperature of the droplet is highest at the bottom of the droplet and lowest at its
 top} \cite{zeng2020}.
\textcolor{black}{As the surface tension decreases with increasing temperature, the surface tension is highest at the top of the water droplet} \cite{sugden1924}.
This causes an off-wall  thermal Marangoni flow \cite{zeng2021pnas}.
The Marangoni flow brings the hot 1-butanol solution with a high water concentration $c_w$ to the cold region above the droplet, where the water solubility is much lower. Therefore the
 dissolved water then precipitates out as tiny water microdroplets.  These then form  a cloud or plume, as clearly seen in
 Fig. \ref{process}a, (II) - (IV).

One wonders on whether gravity effects also play a role. To find out, we estimate the Archimedes number (also called
Grasshoff number), which compares the driving force due to gravity with the viscous force,
$Ar = g R_\mu ^3  \Delta \rho/(\textcolor{black}{\rho_0 \nu^2})$. The density difference between the (heavier) water and the (lighter)
water-saturated 1-butanol is $\Delta \rho$ = 152 \textcolor{black}{kg/m$^3$} and the kinematic
viscosity of 1-butanol at room temperature
$\nu \approx 4  \times  10^{-6}$ \textcolor{black}{m$^2$/s}. This implies that even a very large microdroplet of   $R_\mu$ = 100 $\mu$m
has an Archimedes number of $Ar \approx 0.1$, so that gravity effects do not play a role.
Even more convincingly, gravity effects would point downwards towards the sessile drop, as water is heavier
than 1-butanol, but the plume of water microdroplets in Fig.  \ref{process}a, (II) - (IV) points upwards
(here off-wall), clearly
ruling out gravitational effects.


\begin{figure}[htpb]
\includegraphics[width=1\textwidth]{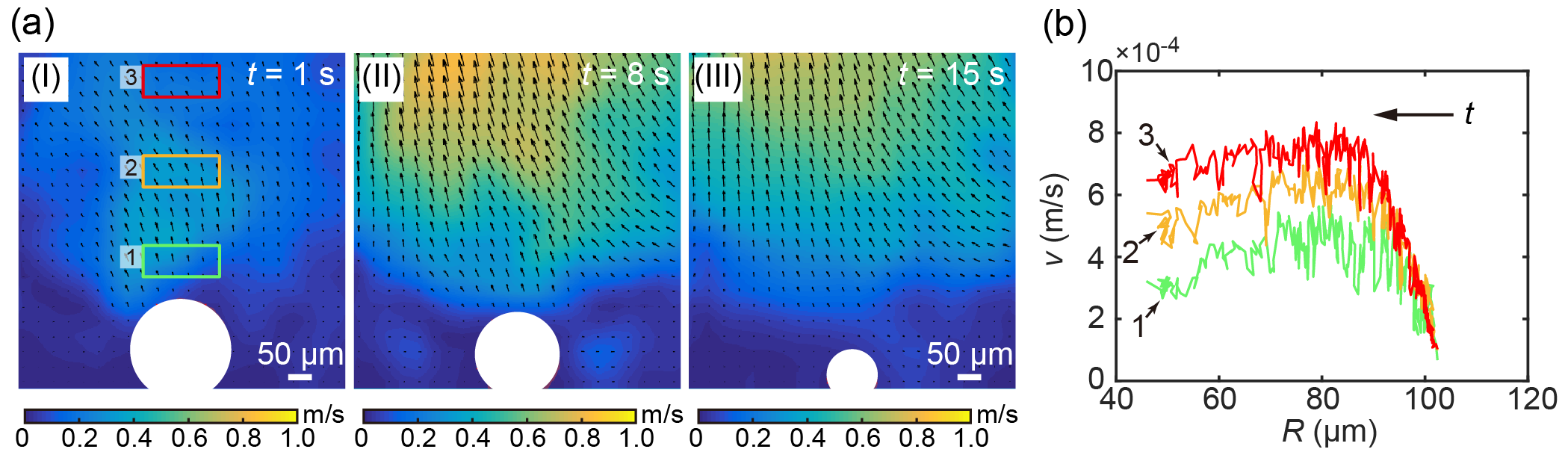}
\caption{(a) Particle image velocimetry of the liquid in the vicinity of a dissolving droplet at different instants of time \textcolor{black}{at a laser power of $P_\ell$ = 40 mW}.
(b) Measured mean liquid velocity as a function of the bubble radius (and thus time) as measured with PIV in the three selected regions 1, 2, and 3, as shown in figure a(I). 
 }
\label{piv}
\end{figure}

To measure the fluid flow caused by the thermal Marangoni effect, we performed particle image velocimetry (PIV) \textcolor{black}{measurements}.
Polystyrene particles (Thermo Scientific, Fluoro-Max Red) with a diameter of 5 $\mu$m and a concentration of 60 $\mu$g/mL were added to the water-saturated 1-butanol solution. The flow velocity profiles near the water droplet at $t$ = 1, 8 and 15 s are shown in Fig. \ref{piv}a. The data clearly reveal that there is an off-wall  flow.
For details of the PIV \textcolor{black}{measurements}, we refer to Movie S2 in the supplementary material.

We extracted the flow velocities $v$ at three selected regions (marked as 1, 2, and 3 in Fig. \ref{piv}a-I) above
 the droplet. The results are shown in Fig. \ref{piv}b. For these three locations,
the flow velocity $v$ near the top of the droplet increases with increasing distance from the droplet. The dependence of the velocity of the polystyrene particles in the boxes outlined in panel \textcolor{black}{(a)} is shown in Fig. \ref{piv}b. Here we take \textcolor{black}{the} velocity $v$ in the selected area marked by the green box as the Marangoni velocity $v_M$. 
As depicted \textcolor{black}{by} the green curve in Fig. \ref{piv}b, $v_M$ increases from 5$\times$10$^{-5}$ m/s
 to 5$\times$10$^{-4}$ m/s during the dissolution process when $R$ decreases from 102 $\mu$m to 90 $\mu$m. After that, $v_M$ gradually decreases to 3$\times$10$^{-4}$ m/s  \textcolor{black}{at a bubble radius of 45 $\mu$m}.

The velocity of the
 thermal Marangoni convection is proportional to the surface tension gradient along the droplet interface, namely $v_M \sim \Delta\sigma \approx 2R(\partial\sigma/\partial T)(\partial T/\partial y)$, where $\partial T/\partial y$ is the temperature gradient in vertical direction. Since the Peclet number (the ratio of thermal convection \textcolor{black}{and} thermal diffusion) $Pe \approx v_M R/\kappa < 0.4$, where
 $\kappa$ = 7.6$\times$10$^{-8}$ m$^2$/s is the thermal diffusivity coefficient in the water-saturated 1-butanol mixture, thermal diffusion plays a more dominant role in the droplet dissolution process than convection. Therefore the plasmonic induced temperature field \textcolor{black}{can, at least after some time, be} considered to be quasi-stationary. Before this quasi-stationary stage, the temperature gradient $\partial T/\partial y$ rapidly increases, resulting in an increase of $v_M$. Subsequently, $\partial T/\partial y$ is stabilized, but the droplet shrinks continuously,  leading eventually to a decrease of $v_M$.

\begin{figure}[htpb]
\includegraphics[width=1\textwidth]{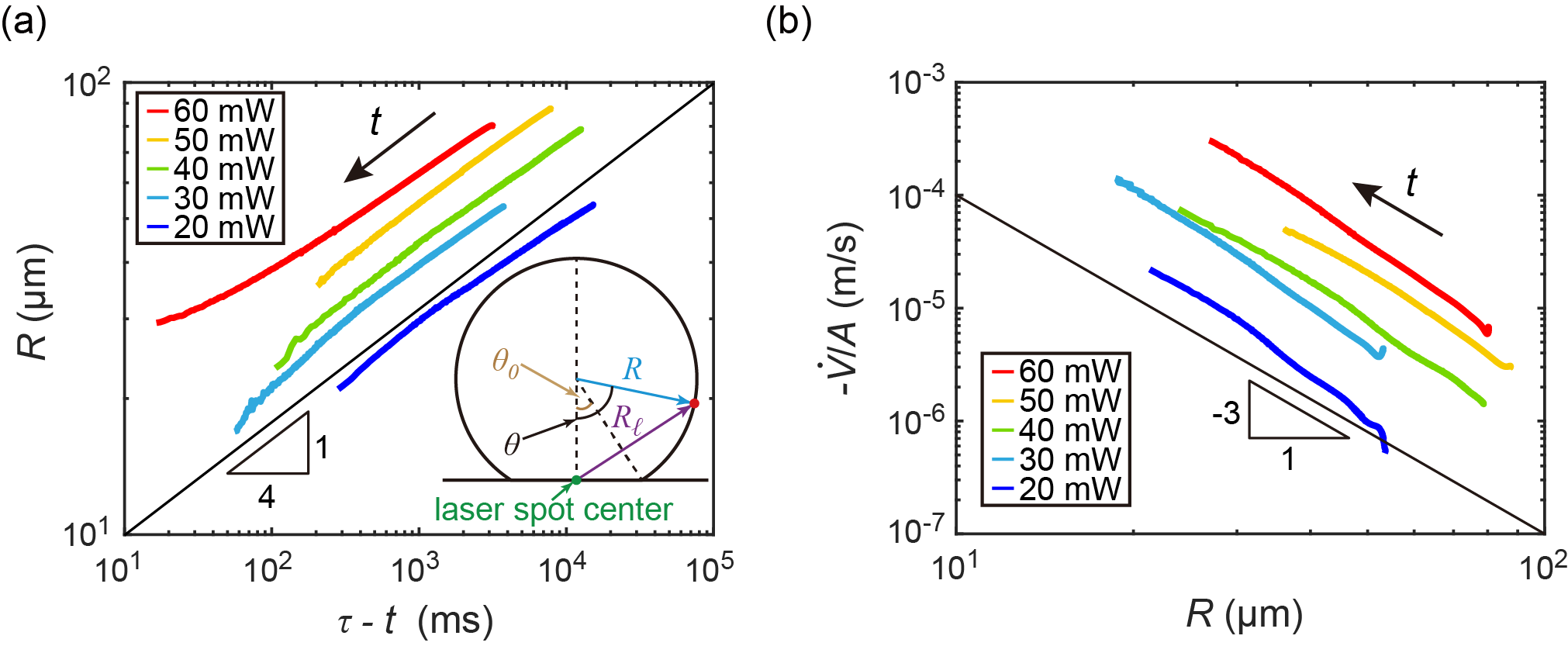}
\caption{(a) Droplet radius $R$ as a function of $\tau - t$ \textcolor{black}{on a} double logarithmic scale. \textcolor{black}{The solid black line refers to an exponent of 1/4.} (b) Rate of volume loss per area $\dot{V}/A$ as a function of droplet radius $R$ at different laser powers $P_\ell$. \textcolor{black}{The solid black line refers to the exponent $-3$.}}
\label{radius}
\end{figure}

To obtain a quantitative analysis of the droplet dissolution dynamics, we systematically conducted a series of experiments at different laser powers.
Figure \ref{radius}a shows the results of $R(t)$ versus $\tau - t$ for different droplets \textcolor{black}{at} different laser powers \textcolor{black}{ranging} from 20 mW  to 60 mW.
The volume loss rate (per area) \textcolor{black}{$\large\dot{V}(t)/A(t)$} versus the droplet radius $R(t)$ is plotted in Fig. \ref{radius}b, \textcolor{black}{where $A(t)\sim (R(t))^2$  is the surface area of the droplet}. From the two plots, \textcolor{black}{we find scaling laws} close to \begin{equation}
R(t)  \sim (\tau - t)^{1/4}\label{main} \end{equation} and $\dot{V}/A \sim R^{-3}$, respectively and consistently.

Equation (\ref{main}) is our main experimental result. We emphasize that this
scaling law (\ref{main}), which implies a droplet lifetime of $\tau \sim R_0^4$,
is pronouncedly
different from $R(t)  \sim (\tau - t)^{1/2}$ as known for pure diffusive dissolution
\cite{lohse2015rmp} and also from $R(t)  \sim (\tau - t)^{4/5}$
known for natural convection dominated dissolution due to  gravity effects \cite{dietrich2016a,chong2020}.
In the next section we will set out to theoretically explain this scaling behavior.

\section{Theoretical explanation of the experimentally found dissolution behavior}

To \textcolor{black}{explain} this new scaling law, a
simplified theoretical model is developed to describe the droplet dissolution process. The mass transfer rate of the droplet \textcolor{black}{is} proportional to the concentration gradient $\left.\partial_{r} c_w\right|_{R}$ at the interface between the water droplet and the surrounding liquid, where $r$ is the radial coordinate from the center of the droplet. The total mass loss rate $\dot{m}$ of the sessile droplet can be obtained by integrating over the water droplet,
\begin{equation}
\dot{m}=\rho \dot{V}=D \int_{\theta_{0}}^{\pi} 2 \pi R^{2} \sin \theta \left.\partial_{r} c_{w}\right|_{R} {\rm d} \theta,
\label{mass_transfer2}
\end{equation}
where $\rho$ is the water density, $D$ is the mass diffusivity coefficient of water in 1-butanol, \textcolor{black}{$\theta_0$ the angle shown in Fig. \ref{radius}a}, $\theta$ is the angle between the droplet radius and the vertical line ($\theta_0$ and $\theta$ are depicted in the inset figure in Fig. \ref{radius}a), \textcolor{black}{and $R$ is the radius of the water droplet}.
Note that we assume \textcolor{black}{that the} water concentration $c_w$ at
 the droplet interface is always in equilibrium. \textcolor{black}{The} water concentration gradient $\left.\partial_{r} c_w\right|_{R}$ of the droplet interface \textcolor{black}{is given by},

\begin{equation}
\left.\partial_{r} c_{w}\right|_{R}=\frac{\partial c_{s}}{\partial T} \left.\partial_{r} T\right|_{R} = 2 a\left(T-T_{0}\right) \left.\partial_{r} T\right|_{R},
\label{Cw_gradient}
\end{equation}
where $c_s$ is the saturation concentration of water in 1-butanol and $a$ = \textcolor{black}{0.0118} kg/(m$^3$K$^2$)
is a known material parameter.
For details, we refer to Fig. \ref{solubility} in Appendix A.
Note that the laser heating can be approximated as point source, since most of the time the laser spot is smaller than the droplet size. In addition, as mentioned earlier, the temperature field can be taken as quasi-stationary, i.e.,
$\partial_t T = 0$, leading to the
Laplace equation
\begin{equation}
\partial_{t} T(r_\ell, t)=\kappa \frac{1}{r_\ell^{2}} \partial_{r_\ell}\left(r_\ell^{2} \partial_{r_\ell} T(r_\ell, t)\right)=0,
\label{laplace}
\end{equation}
where $r_\ell$ is the radial coordinate from the laser spot center and
 $\kappa$ is the thermal diffusivity. By solving Eq. (\ref{laplace})
  at $r_\ell = R_\ell$, where $R_\ell$ is the distance between droplet interface and the laser spot center (Fig. \ref{radius}a), we obtain
\begin{equation}
\left.\partial_{r_\ell} T\right|_{R_\ell}=-\frac{b}{R_{\ell}^{2}},   \qquad     T - T_0= \frac{b}{R_{\ell}},
\label{Tgradient}
\end{equation}
where $b$ is a prefactor.
With $\left.\partial_{r_\ell} T\right|_{R_\ell}$, we can now
obtain the normal component $\left.\partial_{r} T\right|_{R}$ to the spherical droplet surface,
\begin{equation}
\left.\partial_{r} T\right|_{R}=\left.\partial_{r_\ell} T\right|_{R_\ell} \frac{\partial R_{\ell}}{\partial R}=\left.\partial_{r_\ell} T\right|_{R_\ell} f(\theta),
\label{normal}
\end{equation}
where $f(\theta) = \sqrt{\sin ^2 \theta + (\cos \theta_0-\cos \theta)^2}$.

By substituting Eqs. (\ref{Cw_gradient}), (\ref{Tgradient}), and (\ref{normal})
 into Eq.\  (\ref{mass_transfer2}), we obtain
\begin{equation}
\dot{V}=-\frac{4 \pi a b^{2} D}{\rho} \int_{\theta_{0}}^{\pi} \frac{\sin \theta}{f(\theta)^{2}} {\rm d} \theta \frac{1}{R} \sim \frac{1}{R}.
\label{mass_trans_result}
\end{equation}
Eq.\ (\ref{mass_trans_result})
can easily be integrated, resulting in
 $\dot{V}/A \sim R^{-3}$,
  and $R  \sim (\tau - t)^{1/4}$.  These results are
 consistent  with our  experimental results  as seen from Figs.\  \ref{radius}b and a.
 We also obtain  the dissolution time or lifetime $\tau$ of the droplet as
\begin{equation}
\tau = \frac{(2-\cos\theta_0)(1+\cos\theta_0)^2\rho}{16ab^2D\int_{\theta_0}^{\pi} \frac{\sin \theta}{f(\theta)^{2}} {\rm d} \theta} R_0^4 \sim R_0^4.
\label{tau}
\end{equation}

\begin{figure}[htpb]
\includegraphics[width=0.5\textwidth]{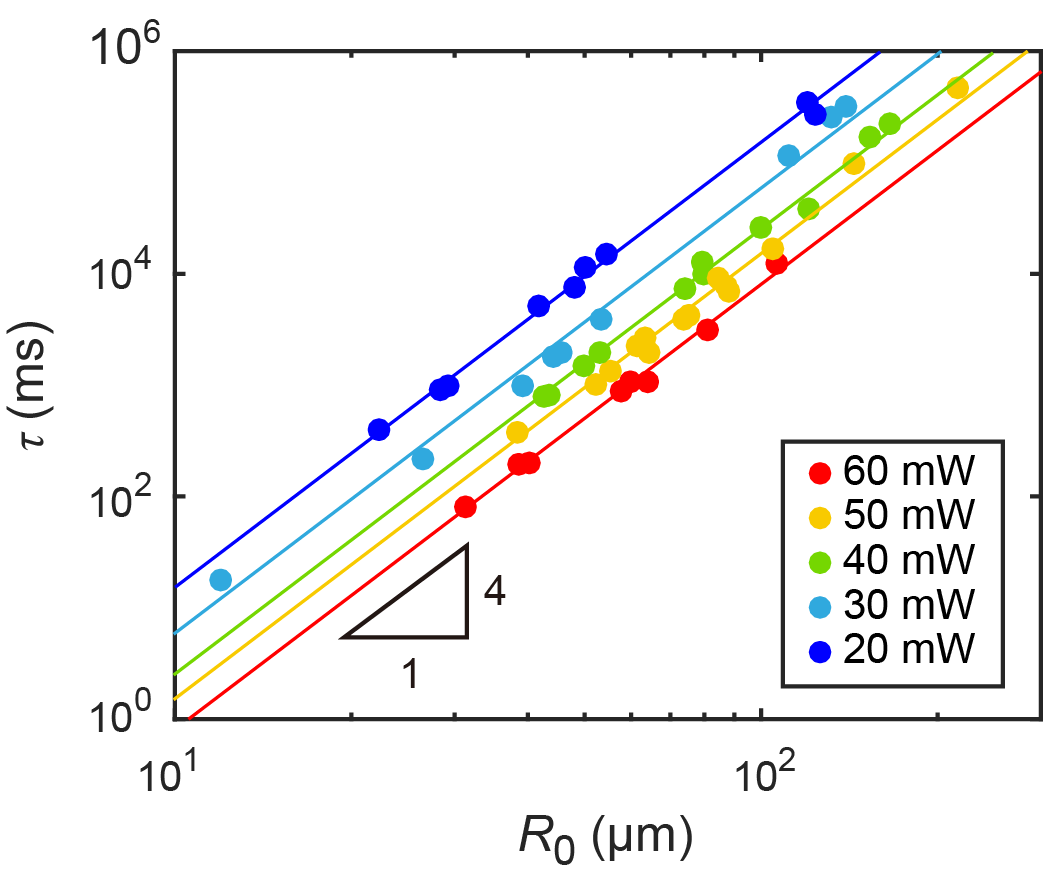}
\caption{Dissolution time $\tau$ of water droplets as a function of the initial
droplet radius $R_0$ at different laser powers $P_\ell$. The droplet lifetime $\tau$ obeys the scaling law of $\tau \sim R_0^4$. The solid lines are drawn to guide the eye.
}
\label{scaling_tau}
\end{figure}

To obtain further evidence for our theoretical model, we measured the lifetime of
water droplets with various initial radii $R_0$
 varying in the  large range of 10 $\mu$m to 200 $\mu$m at different laser powers $P_\ell$.
The results are summarized in Fig.\  \ref{scaling_tau}.
The agreement between the scaling law
Eq.\  (\ref{tau})  and the experimental results is excellent. This supports the validity of
the proposed  scaling  for all considered laser powers and droplet sizes.

\section{Conclusions and outlook}
In summary, we have experimentally and theoretically studied the dissolution of a
 water droplet in a  water-saturated 1-butanol solution.
 The laser irradiation induced plasmonic heating of the liquid around the droplet results in
 (1) the \textcolor{black}{uptake} of water in the 1-butanol and (2) an off-wall  thermal Marangoni flow.
As a result, the hot 1-butanol solution with a relative high water concentration \textcolor{black}{is transported} to the cold region, where the water solubility is lower. This leads to the formation of a plume consisting of tiny water microdroplets.
Most importantly, the thermal gradient and the thermal Marangoni flow around the droplet
affect its dissolution dynamics and make
it very different from isothermal and  purely
diffusive dissolution dynamics or the  dissolution dynamics due to natural convection,
which lead
to droplet lifetimes of $\tau \sim R_0^2$ and $\tau \sim R_0^{5/4}$, respectively. In contrast, here we have
experimentally found that
 the droplet lifetime scales as the fourth power of the initial bubble radius,
$\tau \sim R_0^4$, and have provided a theoretical model  to account for this new dissolution
behavior.

We expect that non-standard droplet or bubble dissolution laws can also hold in other situations where
thermal or solutal Marangoni flows play a role, including those of relevance in applications, such as
in catalysis or electrolysis, where bubbles can evolve at catalytic surfaces or on  electrodes \cite{yang2015,bashkatov2019,yang2018marangoni,lohse2018}.
In these cases  the bubbles are hindering the catalytic or electrolytic processes and an understanding of the
effect of the Marangoni forces and the temperature dependent solubility
may help to enhance the bubble dissolution. Another situation where Marangoni forces may affect the
dissolution behavior are  droplets driven by autochemotaxis
\cite{bekki1992,chen2009self,jin2017pnas,liebchen2018,hokmabad2021}.

\vspace{3mm}\noindent
{\it Acknowledgements:} The authors thank the Dutch Organization for Research (NWO) and the Netherlands Center for Multiscale Catalytic Energy Conversion (MCEC) for financial support. Y.W. appreciates financial support by National Natural Science Foundation of China (Grant No. 51775028 and 52075029). D.L. acknowledges financial support by an ERC Advanced Grant DDD under the project No. 740479 and by NWO-CW. B.Z. thanks the Chinese Scholarship Council (CSC) for financial support.

\section*{Appendix A: Solubility dependence}

\begin{figure}[htpb]
\includegraphics[width=0.5\textwidth]{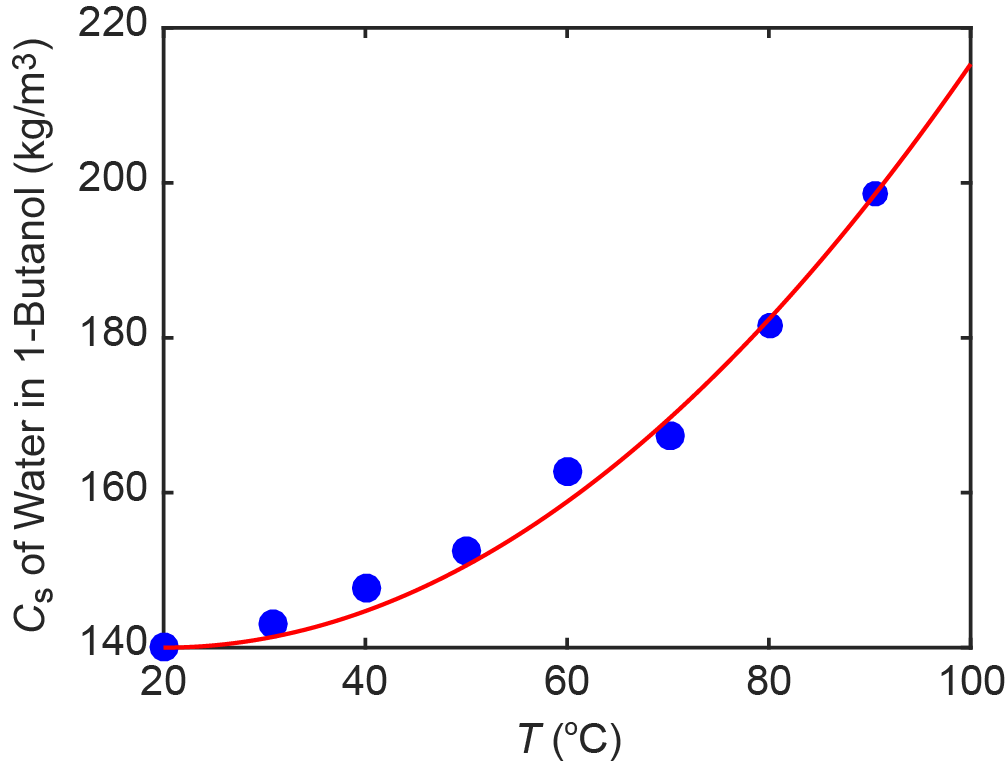}
\caption{ The saturation concentration $c_s$ of water in 1-butanol as a function of the temperature $T$. Data points are from Ref. \cite{Stephenson1986}. The solid curve is the result of a  least-square quadratic fitting of the data points to
Eq.\  (\ref{c_s}).}
\label{solubility}
\end{figure}

The saturation concentration (solubility) $c_s$ of water in 1-butanol as a function of temperature $T$ is shown in Fig. \ref{solubility}. It is found that $c_s$ can be well described by  a quadratic function in the temperature
 $T$,
\begin{equation}
c_s - c_{s0} = a(T - T_0)^2,
\label{c_s}
\end{equation}
where $c_{s0}$ is the saturation concentration of water in 1-butanol at room temperature $T_0$ = 20$^\circ$C, and $a$ = \textcolor{black}{0.0118} kg/(m$^3$K$^2$) is a material constant. By taking the  derivative of Eq. (\ref{c_s})
 with respect to $T$, we obtain
\begin{equation}
\frac {\partial c_s}{\partial T} = 2a(T - T_0).
\label{partial_c_s}
\end{equation}

\bibliographystyle{prsty_withtitle}


\bibliography{nanobubble-and-inkjet-literatur}


\end{document}